\documentclass[aps,prb,twocolumn,preprintnumbers,amsmath,amssymb]{revtex4}
\usepackage{graphicx,xcolor}
\begin{document}
\title{Quantum criticality features in the Co doped MnSi.}

\author{S.~M.~Stishov}
\email{stishovsm@lebedev.ru}
\affiliation{P. N. Lebedev Physical Institute, Leninsky pr., 53, 119991 Moscow, Russia}
\author{A.~E.~Petrova}
\affiliation{P.~N.~Lebedev Physical Institute, Leninsky pr., 53, 119991 Moscow, Russia}
\author{A.~M.~Belemuk}
\affiliation{Institute for High Pressure Physics, Russian Academy of Science, Troitsk 108840, Russia}

\begin{abstract}
The mysterious universal line revealing an independence of spin fluctuation contributions to the heat capacity of (Mn,Co)Si on impurity contents and its nature is discovered. This situation probably declares an invariance of the spin subsystem energy that may provide by the response of itinerant electron system on the volume change at doping.
\end{abstract}
\maketitle
\section{Introduction}
\begin{figure}[htb]
\includegraphics[width=60mm]{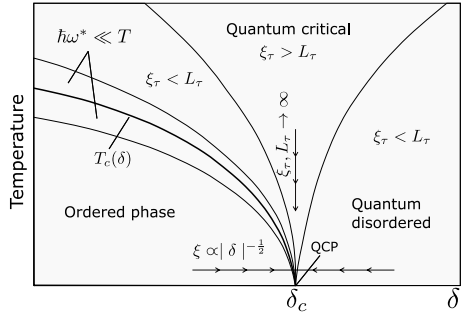}
\caption{\label{fig1} Schematic phase diagram of a substances in the vicinity of a quantum critical point.
$\xi_{\tau}\varpropto \xi^{z}$ is the correlation time, $ L_{\tau}=\hbar/kT$ is time coordinate. 
The lines $ \xi_{\tau}=L_{\tau} $ correspond to the quantum - classical crossover.}
\end{figure}

 The Quantum critical phenomena are expected when a second order phase transition line comes to zero temperature as shown in Fig.~\ref{fig1}~\cite{1,2,3,4,5}.  
As is seen in Fig.~\ref{fig1} a quantum critical point (QCP) occurs at variation of a control parameter $\delta$, which can be pressure, magnetic field or chemical concentration. The QCP is situated at the zero-temperature end of a singular critical trajectory along which the correlation length and correlation time both diverge. In case of metallic magnets, the ratio $C_{p}/T$ also diverge along the critical trajectory that indicates a diverging electron effective mass “dressed” by spin fluctuations.
In the current paper some analysis of the experimental data~\cite{6,7,8} on MnSi phase diagram doped with Co and Fe is given. As was discovered some years ago helical magnet MnSi experiences a quantum phase transition at pressure about 14 kbar~\cite{9,10,11}, but it is still not clear whether this transition is quantum critical or not~\cite{12}]. So it was suggested that a doping MnSi with suitable impurities may shed a new light on this problem.
However, it was found in the recent study of MnSi doped with Co~\cite{6,7} that in this case there is no a singular critical trajectory. Instead, a cloud of the helical fluctuations with a diverging ratio $C_{p}/T$ spreading over a significant range of concentrations and temperatures arise at large concentration of Co close to 0~K. Yet, some specifics of heat capacity behavior in (Mn,Co)Si requires a further consideration.
\section{Experimental data and discussion}
\begin{figure}[htb]
\includegraphics[width=60mm]{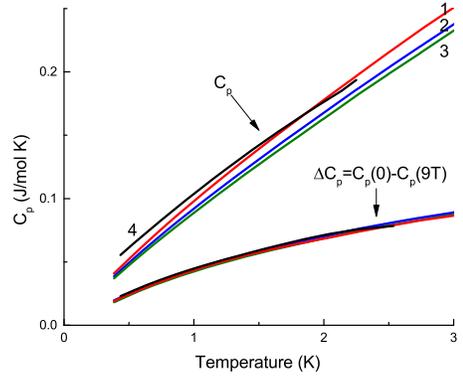}
\caption{\label{fig2} Two sets of heat capacity data. Designations shown in the plot. The data were moderately smoothed for better viewing. (Mn$_{1-x}$Co$_x$)Si: (1, 2, 3)-0.09, 0.063, 0.057, (Mn$_{0.83}$Fe$_{0.17}$)Si: (4)}. 
\end{figure}
Some magnetic and heat capacity measurements were performed to characterize the (Mn,Co)Si samples. All measurements were made by use the Quantum Design PPMS system with the heat capacity and vibrating magnetometer moduli and the He-3 Refrigerator. 
The samples were prepared by arc melting under argon atmosphere and subsequently the triarc Czochralski technique. The electron-probe microanalysis shows that compositions are: Mn$_{0.93}$Co$_{0.057}$Si, Mn$_{0.92}$Co$_{0.063}$Si, Mn$_{0.89}$Co$_{0.09}$Si,  Mn$_{0.83}$Fe$_{0.17}$Si.  The lattice parameters of these compounds obtained by the X-ray diffraction technique are displayed in Fig. 3. 
\begin{figure}[htb]
\includegraphics[width=60mm]{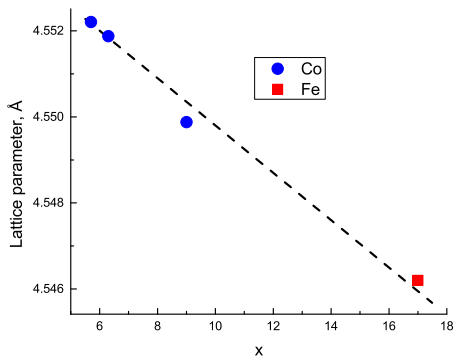}
\caption{\label{fig3}The lattice parameters of (Mn$_{1-x}$Co$_x$)Si: x=0.057,0.063,0.09 and (Mn$_{0.83}$Fe$_{0.17}$)Si Ref.~\cite{6,8} .} 
\end{figure}
\begin{figure}[htb]
\includegraphics[width=60mm]{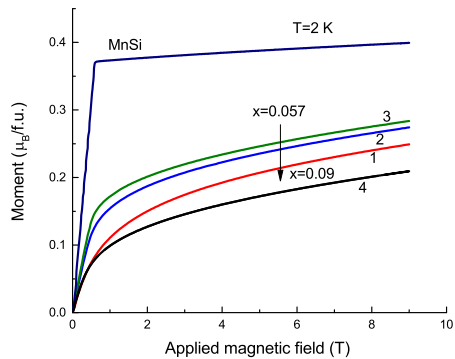}
\caption{\label{fig4} Magnetization curves for (Mn$_{1-x}$Co$_x$)Si:(1, 2, 3)-0.09, 0.063, 0.057,  and Mn$_{0.83}$Fe$_{0.17}$)S: (4) in comparison with one for pure MnSi Ref.~\cite{6,8} .} 
\end{figure}
For comparison purpose the classical Monte Carlo calculations were made to describe the behavior of magnetization of a spin system in an applied magnetic field. We use an approach involving localized spins coupled by the exchange and Dzyloshinski-Moriya (DM) spin-spin interactions. A standard lattice spin Hamiltonian for localized spins in an applied magnetic field reads  \cite{13,14,15}
Upon doping a regular spin is replaced by an effective impurity spin  which is supposed to be a classical spin of unit length similar to the regular one.
The impurity spins are coupled with neighboring regular spins by some modified exchange and DM coupling constants  If two impurity spins happen to occur in neighboring sites the corresponding coupling constants is forced to be zero.
The MC simulation was carried out on a  cubic lattice with periodic boundary conditions using a standard single-site Metropolis algorithm. I We made $10^6$ MC steps per spin (MCS) to equilibrate the system and next $10^6$ MCS (and up to $10^7$ MCS in separate runs) to gain statistics.
From the simulation we directly find the $z$-component (along applied field) of the magnetization.More details on the MC procedure with impurities and the choice of the coupling parameters of the effective model given in Refs. \cite{16,17}
To discuss the mentioned specifics of the heat capacity let us to turn to Fig.~\ref{fig2}, where two sets of the heat capacity curves are shown. One of the sets includes heat capacity data for (Mn,Co)Si~\cite{6} and (Mn,Fe)Si~\cite{8} with appending new experimental data on (Mn,Co)Si below 2~K~\cite{7}. The other set illustrates the difference between heat capacity at zero magnetic field $C_p(0)$ and heat capacity at 9~T ($C_p(9 T)$ for (Mn,Co)Si samples  and the sample of (Mn,Fe)Si. As was mentioned in Ref.~\cite{6} this manipulation implies a subtraction of some of background contribution, including phonon and electron ones to the heat capacity leaving its fluctuation part intact. As is seen in 
\begin{figure}[htb]
\includegraphics[width=50mm]{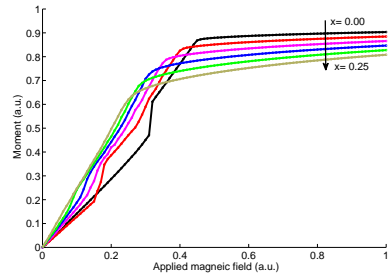}
\caption{\label{fig5} Magnetization curves for the classical chiral spin system with isomorphic impurities. Parameter $x$ designate the fraction of impurity spins in the regular lattice. Some irregularities are probably result of metastability phenomena.  }. 
\end{figure}
\begin{figure}[htb]
\includegraphics[width=60mm]{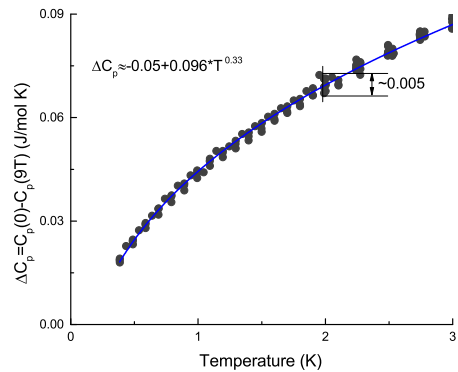}
\caption{\label{fig6} Fitting of the low temperature part of differential heat capacity of all four experimental sets to the power function, the value of the power exponent shown in the plot. The constant 0.05 in the fitting corresponds to 0.15 K at 
$\Delta C_{p}=0$ that is some revelation of systematic errors at the heat capacity measurements.  } 
\end{figure}
Fig.~\ref{fig2} the mentioned procedure results in the puzzling universal line exposing an independence of the fluctuation contributions to the heat capacity on impurity contents and its nature. This situation suggests an invariance of this contribution  despite the all changes caused by doping. This bold guess seems to contradict the observations. Indeed, the fluctuations under discussions are spin fluctuations an they should be connected with a general spin number in a system.  The saturation magnetization data clearly indicate a progressive magnetic moment (spins?) lost on doping Fig.~\ref{fig4}. The latter is supported by the Monte Carlo calculations of the saturation magnetization of the classical chiral spin system with isomorphic impurities (Fig.~\ref{fig5}). However, the volume decrease on doping (see Fig.~\ref{fig3}) may keep the energy of  spin subsystem almost compensated for the spin lost and therefore makes the spin fluctuation abundance not quite sensitive to the spin replacements.      
\begin{figure}[htb]
\includegraphics[width=60mm]{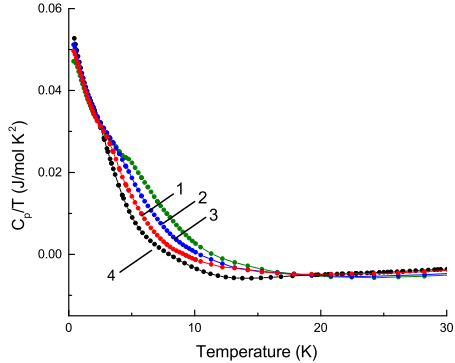}
\caption{\label{fig7} The ratio (C$_{p}$-C$_{p}$(H))/T as a function of temperature.Mn$_{1-x}$Co$_x$)Si: (1, 2, 3)-0.09, 0.063, 0.057, (Mn$_{0.83}$Fe$_{0.17}$)Si: (4)} 
\end{figure}
\begin{figure}[htb]
\includegraphics[width=60mm]{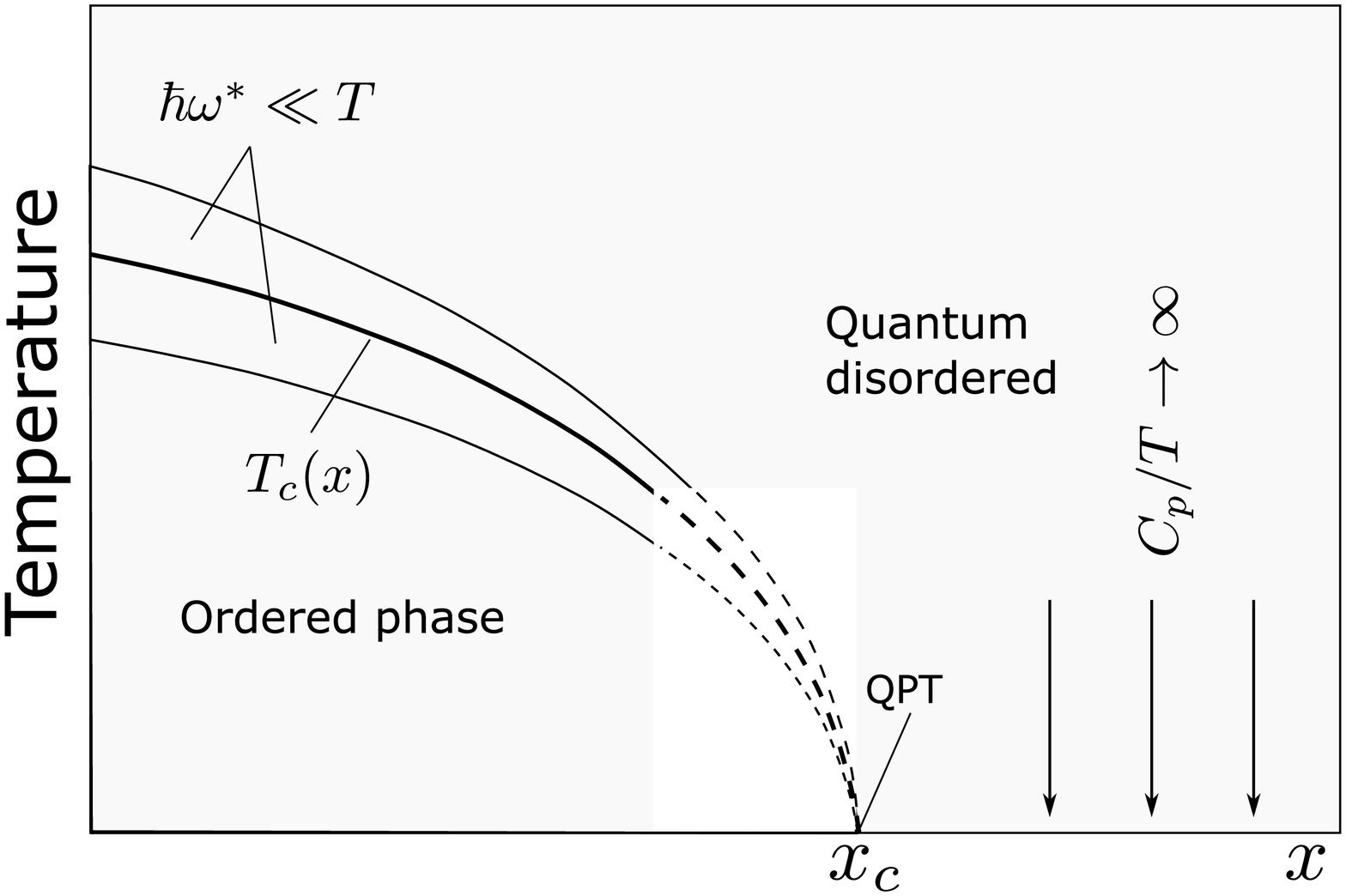}
\caption{\label{fig8}Proposed schematic phase diagram of Mn$_{1-x}$Co$_{x}$Si. Dashed lines indicate that at large impurity concentration the phase transition is strongly smeared out. The phase transition at \textit {X}$_{c}$ can be quantum critical point or quantum first order phase transition.     } 
\end{figure}
\section{Conclusion}
As is shown in Fig.~\ref{fig6}(see also Fig.~\ref{fig2}) the fluctuation contribution to the heat capacity can be described with a single power function of temperature for all four data sets:  $\Delta C_{p}\sim T^{0.335}$, . Correspondingly, the ratio $\Delta C_{p}/T$ diverge as $T^{-0.665}$ Fig.~\ref{fig7}, therefore, confirming that just the spin fluctuations are responsible for the divergence of $C_{p}/T$. In this connection we are reminded that for metallic systems at $T\rightarrow 0$, $C_{p} = \gamma T$ or $C_{p}/T = \gamma$, where $\gamma \sim m^{*}$, $m^{*}$is the effective electron mass. So,the effective electron mass diverges at $T\rightarrow 0$, being “dressed” by the spin fluctuations. The mysterious universal line revealing an independence of spin fluctuation contributions to the heat capacity of (Mn,Co)Si on impurity contents and its nature is discovered. This situation probably declares an invariance of the spin subsystem energy that may provide by the response of itinerant electron system on the volume change at doping.
The current data are revealing once again that a singular quantum critical point does not exist in the system under study. In its place, one can see some sort of a quantum critical cloud covering a significant range of dopant concentrations. 
This finding suggests a phase diagram for metal magnet with doping as a control parameter, presented in Fig.~\ref{fig8}, where as a difference to the diagram in Fig.~\ref{fig1} there is no a singular quantum critical trajectory. One can see here a parallel to the $T-P$ phase diagram of pure MnSi with pressure as a control parameter, where an extended paramagnetic region with specific properties beyond the phase transition point was found in Ref.~\cite{18,19}.
	

\begin{thebibliography}{99}
\bibitem{1} S. Sachdev, Quantum Phase Transitions, Cambridge: Cambridge
Univ. Press (1999)
\bibitem{2} M.A. Continentino, Quantum Scaling in Many-Body Systems,
Singapore: World Scientific (2001)
\bibitem{3} S.L. Sondhi  et al. Rev. Mod. Phys. \textbf{69}, 315 (1997)
\bibitem{4} T. Vojta Ann. Phys. (Leipzig)\textbf{9}, 403 (2000)
\bibitem{5} C.M. Varma , Z. Nussinov, W.van Saarloos,  Phys. Rep.\textbf{361}, 267 (2002)
\bibitem{6} A.E. Petrova, S.Y. Gavrilkin, G.V. Rybalchenko, D. Menzel,I. P. Zibrov, and S. M. Stishov, Phys. Rev. B \textbf{103}, L180401 (2021).
\bibitem{7} A.E. Petrova, S.Yu. Gavrilkin, and A.Yu. Tsvetkov, Dirk Menzel and Julius Grefe, S. Khasanov, S.M. Stishov, Phys. Rev.B\textbf{106} 014406 (2022)
\bibitem{8} A.E. Petrova, S. Yu. Gavrilkin, D. Menzel, and S.M. Stishov, Phys. Rev. B \textbf{100}, 094403 (2019).
\bibitem{9} J.D. Thompson, Z. Fisk, and G. G. Lonzarich, Phys. B
(Amsterdam, Neth.) \textbf{161}, 317 (1989).
\bibitem{10} C. Peiderer, G. J. McMullan, and G. G. Lonzarich, Phys.B (Amsterdam, Neth.)\textbf{206}, 847 (1995).
\bibitem{11} C. Thessieu, J. Flouquet, G. Lapertot, A. N. Stepanov, and D. Jaccard, Solid State Commun. \textbf{95}, 707 (1995).
\bibitem{12} S.M. Stishov, A.E. Petrova, Physics - Uspekhi,\textbf{60(12)}, 1268 (2017)
\bibitem{13} S. D. Yi, S. Onoda, N. Nagaosa, J. Hoon Han, Phys. Rev. B {\textbf 80}, 054416 (2009).
\bibitem{14}	 A. Hamann, D. Lamago, Th. Wolf, H. v. L\"{o}hneysen and D. Reznik, Phys. Rev. Lett. {\bf 107}, 037207 (2011).
\bibitem{15} S. Buhrandt and L. Fritz, Phys. Rev. B \textbf{88}, 195137 (2013).
\bibitem{16} A. M. Belemuk, S. M. Stishov, Phys. Rev. B \textbf{104}, 064404 (2021)
\bibitem{17} A.M. Belemuk, Solid State Commun. \textbf{351}, 114787 (2022). 
\bibitem{18} C. Pfleiderer, S.R. Julian, G.G. Lonzarich, Nature \textbf{414}, 427 (2001)
\bibitem{19} C. Pfleiderer, D. Reznik, L. Pintschovius, H. v. L\"{o}hneysen, M. Garst, A. Rosch, Nature \textbf{427}, 227 (2004)
\end{thebibliography}
\end{document}